\documentstyle[12pt,psfig]{article}
\begin{document}

\title{$\rho/\omega$ properties from dilepton spectra in $p A$
reactions at 12 GeV
\thanks{Supported by Forschungszentrum J\"ulich}}
\author{E. L. Bratkovskaya  \\[2mm]
{\normalsize Institut f\"{u}r Theoretische Physik, Universit\"{a}t Giessen}\\
{\normalsize 35392 Giessen, Germany}}
\date{ }
\maketitle

\begin{abstract}
The dilepton production from $pC$ and $pCu$ collisions at
$T_{lab}=12$~GeV is calculated using the semi-classical BUU transport
model, that includes the off-shell propagation of vector mesons
nonperturbatively and calculates the width of the vector mesons
dynamically.  It is found that the collisional broadening of the vector
meson width and dropping vector meson masses lead to a small enhancement
of the dilepton yield below the $\omega$ meson mass pole in $p Cu$ collisions
compared to $p C$, which is, however, not sufficient to explain the
enhanced production of dileptons as observed in the experiment by
the KEK-PS E325 collaboration.  It is argued that such in-medium
effects are expected to be small due to  kinematical reasons --  the
dominant part of vector mesons radiate dileptons and decay outside the
nucleus due to the finite formation time for $\rho$ and $\omega$
mesons, which is large by Lorentz covariance since  the produced vector
mesons move with high velocity relative to the target nucleus.
\end{abstract}

\vspace{0.3cm}\noindent


\vspace*{5mm}

The modification of hadron properties in  nuclear matter is of
fundamental interest.  Dilepton data from heavy-ion experiments at
SPS energies \cite{CERES,HELIOS} have provided  first experimental
evidence for a change of the vector meson properties in the nuclear medium,
however, the heavy-ion data can be interpreted within different
scenarios of in-medium modifications, i.e.  by the dropping mass
scenario or the collisional broadening approach
(cf. the reviews \cite{Gerry,CBRep98,Rapp00} and Refs. therein).

Since in heavy-ion experiments the nuclear matter is probed at
different densities and temperatures within the complex dynamical
evolution, it is very useful to have  independent experimental
information from photon-nucleus, pion-nucleus or proton-nucleus
reactions, where the properties of vector mesons are probed at normal
nuclear density or below.
Such information has been provided recently by the KEK-PS E325
collaboration, that has measured the dilepton spectrum in the mid-rapidity
region from $p C$ and $p Cu$ collisions at 12 GeV \cite{KEK}. A
significant difference in the mass spectra below the $\omega$ meson
mass pole has been observed between $p C$ and $p Cu$ reactions. This
difference has been interpreted as a modification of the vector meson
properties in nuclear matter \cite{KEK}.

In this work the question of in-medium modifications of the $\rho,
\omega$ meson properties is studied within the semi-classical BUU
transport approach of Refs. \cite{Effe99gam,EffePhD}. This model
is based on the resonance concept of nucleon-nucleon and
meson-nucleon interactions at low invariant energy $\sqrt{s} \ $
\cite{TeisZP97} by adopting all resonance parameters from the
Manley analysis \cite{Manley}. The high energy collisions
(multiparticle production) -- above $\sqrt{s}$ = 2.6~GeV for
baryon-baryon collisions and $\sqrt{s}$ = 2.2~GeV for meson-baryon
collisions -- are described by  the LUND string fragmentation
model FRITIOF \cite{FRITIOF} with a formation time $\tau_F=0.8$
fm/c. This aspect is similar to that used in the HSD
(Hadron-String-Dynamics) approach \cite{CBRep98,Ehehalt} and the
UrQMD model \cite{Bass}.
The threshold for string production of $\sqrt{s}$ = 2.6~GeV for
baryon-baryon production is not a sensitive parameter of the model
since the inclusive meson production cross sections are modeled to
be the same in the resonance and string model for 2.4 $\leq
\sqrt{s} \leq 3$ GeV. Furthermore, above  $\sim \sqrt{s}$ = 2.2
GeV there are no longer distinct resonance excitations in
meson-baryon collisions such that the string formation model is a
suitable way to describe the continuum excitations of the baryon.
 This combined resonance-string approach
allows to calculate particle production in baryon-baryon and
meson-baryon collisions from low to high energies. By input the
vacuum cross sections for $NN$ and $\pi N$ are reproduced with
good accuracy as demonstrated in Refs. \cite{CBRep98,EffePhD}. At
the energy of 12 GeV considered here essentially the string
formation and decay scheme is of relevance with respect to vector
mesons, which is well tested by experiment. Cross sections
involving baryon resonances might be disputed due to the lack of
experimental data, but the Manley analysis of the experimental
data \cite{Manley} provides a full determination of the baryon
resonance parameters as well as their branching to mesons.
In the present case the low energy resonance dynamics
play a minor role for the dilepton spectra anyway (see below).

The collisional dynamics for proton-nucleus reactions, furthermore, is
described by the coupled-channel BUU transport approach
\cite{Effe99gam,EffePhD} that is based on the same elementary cross
sections.  Furthermore, the model has been extended to include the
off-shell propagation of vector mesons nonperturbatively
\cite{Brat01pA} and to calculate the width of the vector mesons
dynamically, which is consistent with the vector meson
production/absorption amplitudes (or probabilities) \cite{Effe99gam}
(see below).

The dilepton production within the resonance-string model is
treated (as described in Ref.~\cite{Brat01pA}) via the production
of resonances $R$ (baryonic or mesonic) in baryon-baryon,
meson-baryon or meson-meson collisions, which can couple to
dileptons directly or via a further decay into mesons or baryons.
The electromagnetic part of all conventional dilepton sources  --
$\pi^0, \eta, \omega$ and $\Delta$ Dalitz decays, direct decay of
vector mesons $\rho, \omega$ -- are treated in the same way as
described in detail in Ref.~\cite{BCM00SIS}, where dilepton
production in elementary $pp$ and $pd$ reactions has been studied
and compared to the available data.

The treatment of the off-shell propagation of vector mesons is
implemented as in Ref.~\cite{Brat01pA} following the general off-shell
transport dynamics from Refs. \cite{Cass_off}.
Note, that in present analysis the nucleons are propagated in the quasi
particle approximation with respect to an optical potential
\cite{CBJ00}, which becomes negligible at high energies.
In the approach \cite{Cass_off} the equations of motion
for the 'off-shell' test particles are extended by terms involving
their dynamical width $\Gamma(X,P)$ that is related to their
retarded self energy $\Sigma^{ret}(X,P)$ by $\Gamma(X,P)= -  Im
\Sigma^{ret}(X,P)/P_0$, where $X,P$ denote the actual space-time
and 4-momentum coordinates. As shown in Refs. \cite{Cass_off} this
leads to a dynamical spectral function for the meson as
\begin{equation}
\label{spec} A(M;X,P) = \frac{2}{\pi}\frac{M^2
\Gamma(X,P)}{(M^2-M_0^2-Re \Sigma^{ret}(X,P))^2 + (M
\Gamma(X,P))^2}
\end{equation}
where the meson mass pole is also shifted by $Re \Sigma^{ret}(X,P)$.
Note, that nonrelativistic complex optical potentials are obtained from
the retarded self energies by dividing $\Sigma^{ret}(X,P)$ by twice the
energy $P_0$. On the other hand, $\Gamma(X,P)$ can be determined by the
in-medium meson decay and collision rates locally from the transport
calculation, i.e. for fixed $X$ and $P$. In principle, $Re \Sigma^{ret}$
and $Im \Sigma^{ret}$ are related to each other by
in-medium dispersion relations, however, in the actual transport
realization they are modeled independently to explore the sensitivity
to the dilepton spectra.
It should be noted that the modeling of the mass shift $Re \Sigma^{ret}$
and the total width $\Gamma$ is constrained to some extent by QCD sum
rules for the vector meson spectral function. As demonstrated in Ref.
\cite{Leupold} the sum rules require a significant shift of the
$\rho$-meson pole mass at nuclear matter density $\rho_0$ if a narrow
$\rho$ spectral function is involved. Furthermore, in case of a broad
spectral function for the $\rho$-meson some spectral strength is shifted
to lower invariant mass such that QCD sum rules can be fulfilled for
lower shift $Re \Sigma^{ret}$, but larger width $\Gamma$. However,
the uncertainties in the QCD condensates do not  reliably constrain
the shape of the spectral function, i.e. its pole shift and width.

In Ref. \cite{Brat01pA} the model described above has been applied to
study the in-medium modification of vector meson properties in dilepton
production in $p A$ collisions at SIS (1--4 GeV) energies.  The effects
of a collisional broadening of the vector meson width (in line with
Refs. \cite{GKC97}) and dropping vector meson masses (according to the
Hatsuda and Lee \cite{H&L92} or Brown/Rho scaling \cite{BrownRho}) have
been investigated in Ref.  \cite{Brat01pA}, too.  It has been found
that the collisional broadening + 'dropping mass' scenario leads to an
enhancement of the dilepton yield in the invariant mass range $0.5 \leq
M \leq 0.75$ GeV, which is most pronounced for heavy systems (up to a
factor 2 for $p + Pb$ at 3--4 GeV).

In this work the same transport model -- including collisional
broadening of the vector meson width (for $\Gamma(X,P)$) plus the
'dropping mass' scenario ($Re \Sigma^{ret} \neq 0$) -- is applied
for dilepton production in $p C$ and $p Cu$ reactions at
$T_{lab}=12$~GeV as measured by the KEK-PS E325 collaboration
\cite{KEK}.

\begin{figure}[h]
\centerline{\psfig{figure=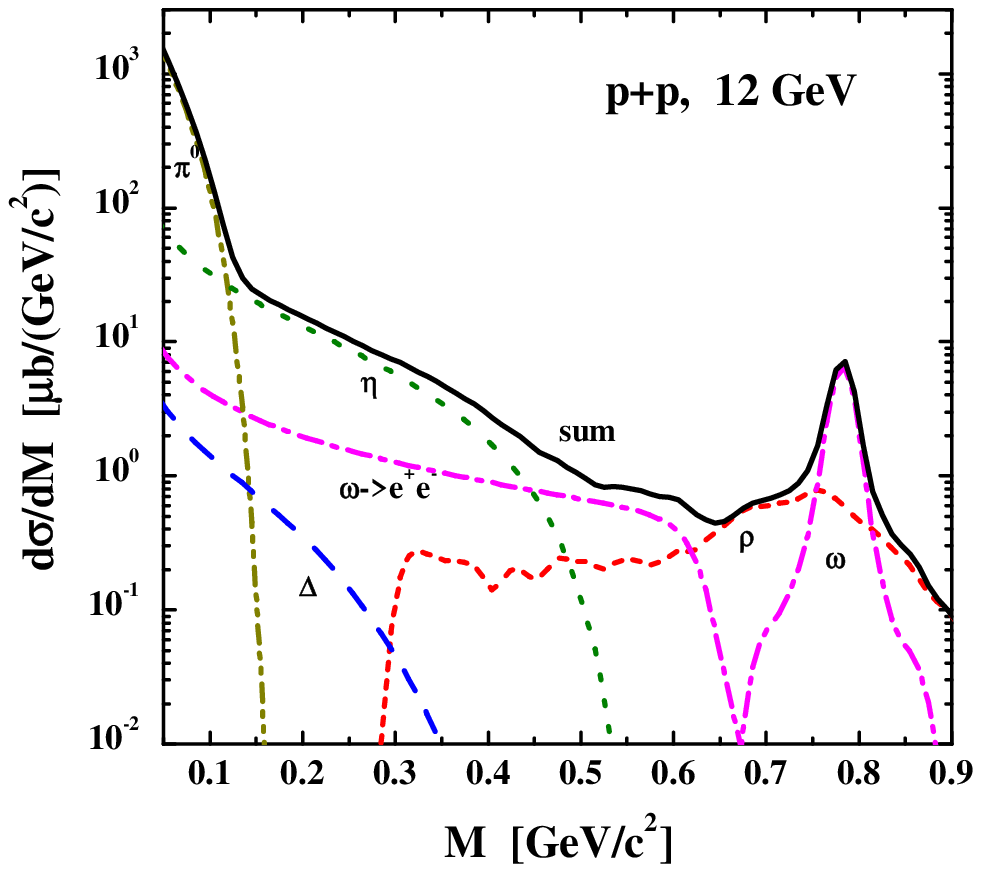,width=10cm}}
\vspace*{0.5cm}
\caption{
The calculated dilepton invariant mass spectra $d\sigma/dM$ for $p + p$
collisions at 12 GeV (including a mass resolution of 9.6 MeV).
The thin lines indicate the individual contributions from the different
production channels; {\it i.e.}~ starting from low $M$:
Dalitz decay $\pi^0 \to \gamma e^+ e^-$ (dot-dot-dashed line),
$\eta \to \gamma e^+ e^-$ (dotted line),
$\Delta \to N e^+ e^-$ (dashed line),
$\omega \to \pi^0 e^+ e^-$ (dot-dashed line),
for $M \approx $ 0.7 GeV: $\omega \to e^+e^-$ (dot-dashed line),
$\rho^0 \to e^+e^-$ (short dashed line).
The full solid line represents the sum of all sources considered here.}
\label{Fig_pp}
\end{figure}

Before showing the results for $p A$ reactions it is worth to
present the 'input' of the calculations, i.e. the calculated
dilepton invariant mass spectra $d\sigma/dM$ for $p + p$
collisions at 12 GeV (Fig. \ref{Fig_pp}).  Note, that in Fig.
\ref{Fig_pp} (as well as in all further figures) a mass resolution
of $\Delta M=9.6$ MeV has been included that corresponds to the
resolution of Ref.~\cite{KEK} for the $\omega$ peak region.  The
thin lines indicate the individual contributions from the
different production channels; {\it i.e.}~ starting from low $M$:
Dalitz decay $\pi^0 \to \gamma e^+ e^-$ (dot-dot-dashed line),
$\eta \to \gamma e^+ e^-$ (dotted line), $\Delta \to N e^+ e^-$
(dashed line), $\omega \to \pi^0 e^+ e^-$ (dot-dashed line), for
$M \approx $ 0.7 GeV: $\omega \to e^+e^-$ (dot-dashed line),
$\rho^0 \to e^+e^-$ (short dashed line).  The full solid line
represents the sum of all sources considered here.  As seen from
Fig. \ref{Fig_pp}, the $\rho$ and $\omega$ decay channels give the
dominant contributions at invariant masses $0.6 \le M \le 0.75$
GeV, where an excess of the dilepton yield in $p Cu$ relative to
$p C$ reactions was observed \cite{KEK}.
In this context it has to be pointed out that the inclusive yield of
$\eta, \rho$ and $\omega$ mesons from the transport approach for $pp$
reactions is consistent with the available experimental data in this
energy regime on the 30\% level \cite{CBRep98}.

\begin{figure}[ht]
\centerline{\psfig{figure=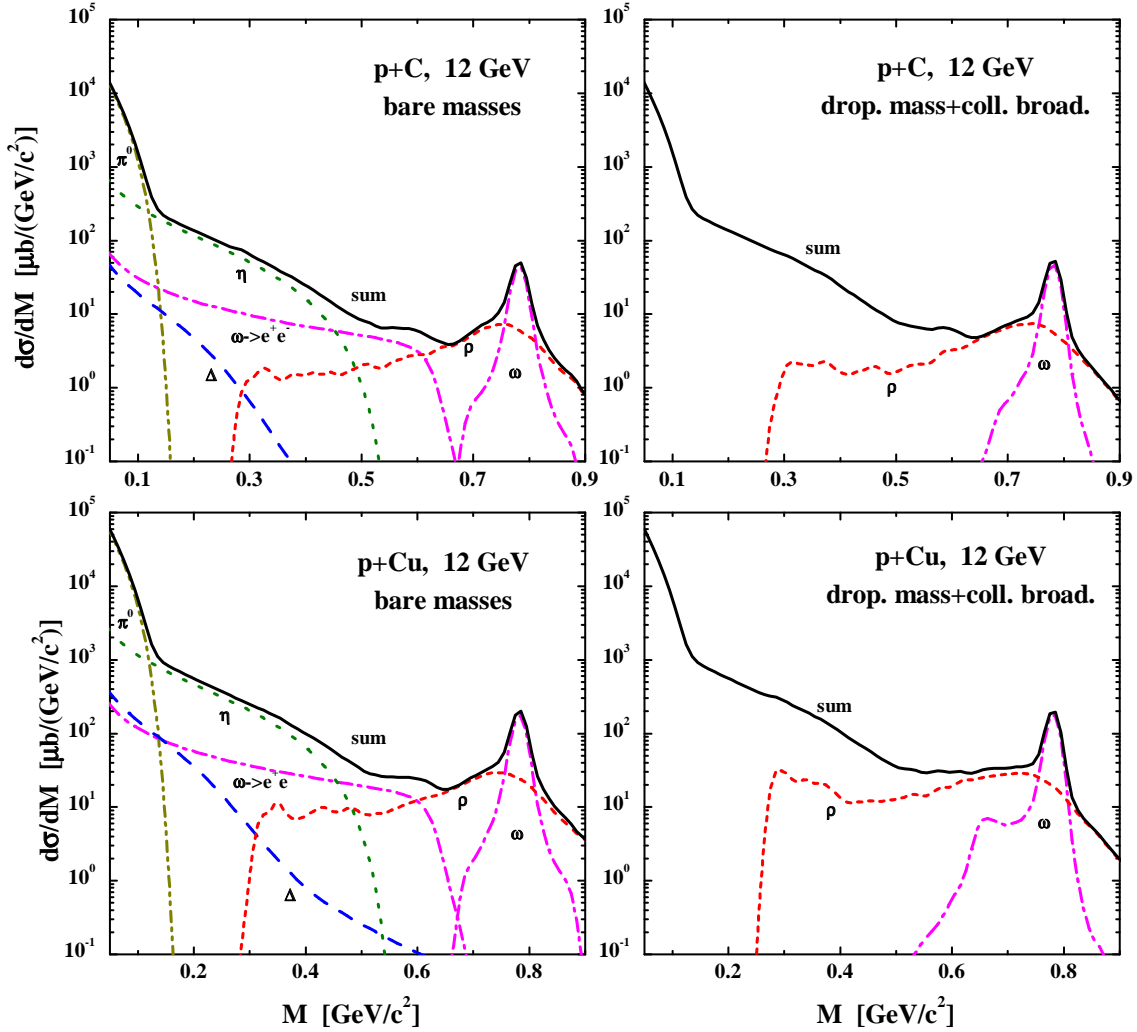,width=15cm}}
\vspace*{0.5cm}
\caption{
The calculated dilepton invariant mass spectra $d\sigma/dM$ for $p + C$
(upper part) and $p + Cu$ collisions (lower part) at 12 GeV (including
a mass resolution of 9.6 MeV) without in-medium
modifications (bare masses; left part), and applying the collisional
broadening + dropping mass scenario (right part).
The assignment of the individual lines is the same as in
Fig. \protect\ref{Fig_pp}.}
\label{Fig_bm}
\end{figure}

In Fig. \ref{Fig_bm} the calculated dilepton invariant mass spectra
$d\sigma/dM$ are displayed for $p + C$ (upper part) and $p + Cu$
collisions (lower part) at 12 GeV (including a mass resolution
of 9.6 MeV) without in-medium modifications (bare masses;
left part), and applying the collisional broadening + dropping mass
scenario  (right part).  The assignment of the individual lines is the
same as in Fig.\ref{Fig_pp}.
Comparing left and right panels one can see that the modification
of the dilepton spectrum by including collisional broadening and
dropping vector meson masses is rather moderate for the light $C$-target
(upper part of Fig.~\ref{Fig_bm}).
For the heavy $Cu$ target (lower part of Fig.~\ref{Fig_bm}) the shift of
the $\rho$ yield to lower invariant masses as well as the secondary
$\omega$ peak (at $\simeq 0.65$ GeV) is seen due to the dropping of
vector meson masses.  However, the total enhancement of the dilepton yield
is quite small, i.e. less than a factor of 1.5 for $0.6 \le M \le 0.75$ GeV.

The direct comparison of the model calculations to the
experimental data requires a precise knowledge of the experimental
acceptance since in the KEK-PS E325 experiment the low part of the
dilepton spectra was significantly suppressed by the geometry of
the detectors and nontrivial electronic cuts \cite{Ozawa_pr}.
Furthermore, the combinatorial background -- from a mixing of
$e^-\pi^+$ and $e^+\pi^-$ pairs -- was not subtracted from the data
\cite{KEK}.  Since the full experimental acceptance is not available, a
normalization procedure -- following the analysis in Ref.  \cite{KEK}
(described below) -- has been applied to compare the calculated spectra
to the experimental data \cite{KEK}.

\begin{figure}[ht]
\centerline{\psfig{figure=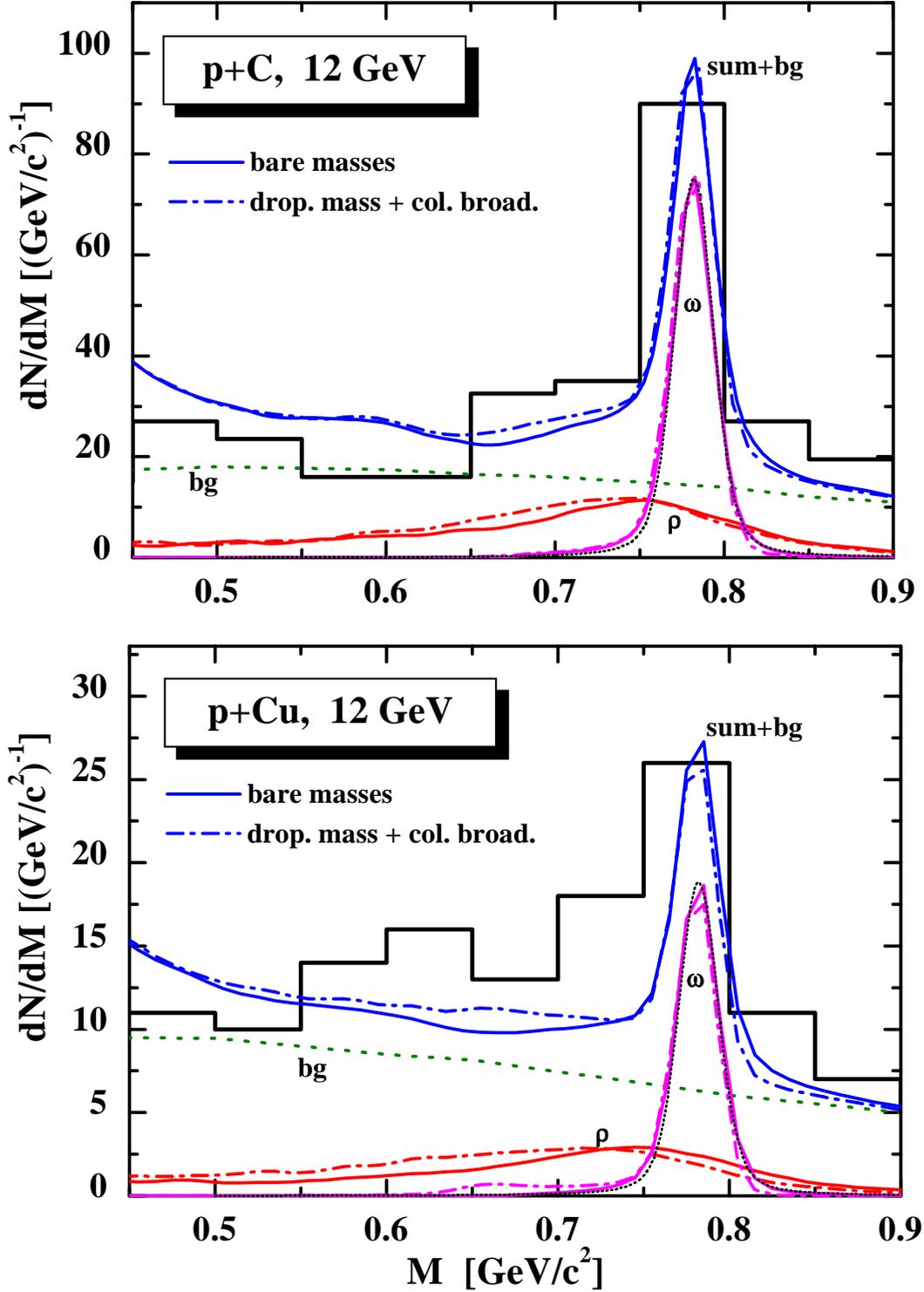,width=14cm}}
\caption{ The calculated dilepton invariant mass spectra $dN/dM$
for $p + C$ (upper part) and $p + Cu$ collisions (lower part)  at
12 GeV (including an experimental mass resolution of 9.6 MeV)
without in-medium modifications (bare masses, solid lines), and
applying the collisional broadening + dropping mass scenario
(dot-dashed lines) in comparison to the experimental data
\protect\cite{KEK} (histograms). The dotted lines show the
combinatorial background from Ref. \protect\cite{KEK} while the
short dotted lines represent the fit to the $\omega$ decay
spectrum from Ref. \protect\cite{KEK}.} \label{Fig_exp}
\end{figure}

The experimental data -- solid histogram in Fig.~\ref{Fig_exp} --
have been fitted in Ref. \cite{KEK} by the  sum of the amplitudes
of $\rho/\omega \to e^+e^-$, $\phi \to e^+e^-$, $\eta \to e^+e^-$
and the combinatorial background (dotted lines in
Fig.~\ref{Fig_exp}, denoted as 'bg') using a four parameter fit.
The mass shape of the $\rho, \omega$ and $\phi$ mesons has been
taken of the Breit-Wigner form with the natural widths 150, 8.41
and 4.43 MeV (cf. Fig. 3 from Ref. \cite{KEK}). The normalization
of the transport calculations is done in a such way that the
integral over the $\omega\to e^+e^-$ contribution for the bare
mass cases (i.e. without in-medium effects) is the same as for the
experimentally fitted $\omega$ contribution (short dotted lines),
that also employs free spectral functions for the $\rho$ and
$\omega$. Note, that the physical conclusions drawn in \cite{KEK}
relate to the same fitted spectra in case of pure vacuum decays.
The same normalization factor is then used for the calculations
with collisional broadening + dropping mass scenario
which allows to study separately the change of the dilepton
spectra when including in-medium effects.
The combinatorial background (taken from Ref. \cite{KEK}) is added
finally to the calculated spectra.

It is worth to point out, that an application of the acceptance
cuts as described in Ref. \cite{KEK}, i.e. the rapidity cut $0.6
\le y \le 2.2$, the transverse momentum cut  $p_T\le 1.5$ GeV, the
cut in the opening angle of the $e^+e^-$ pairs  $50^0 \le
\Theta_{open} \le 150^0$, and the restriction that the electron
and positron are detected in different detector arms (cf. Fig. 1
in Ref. \cite{KEK}),  leads to a suppression of the dilepton yield
by about of a factor of 9 for $M\ge 0.5$ GeV and does not change
the shape of the spectra. At low $M$ these cuts suppress the
dilepton spectra more effective, however,  further 'electronic'
acceptance cuts must be applied to reproduce exactly the shape of
the measured spectra. Due to this reason only the $M$-interval $0.45
\le M\le 0.9$ GeV is considered, where an excess of the dilepton yield
has been observed, to reduce the sensitivity to unknown cuts in the
lower mass regime.

The comparison of the calculated results with the experimental data
(histograms) is shown in Fig.~\ref{Fig_exp} for $p + C$ (upper part)
and $p + Cu$ collisions (lower part). The solid lines indicate the
calculations without in-medium modifications (bare masses), the
dot-dashed lines correspond to a calculation including the collisional
broadening + dropping mass scenario. The individual contributions are
indicated as '$\rho$', '$\omega$'; the lines marked as 'sum+bg' show
the sum of the dilepton contributions and the combinatorial background
(dotted lines 'bg') as taken from Ref. \cite{KEK}.

Note, that the $\rho$-contribution in Fig. \ref{Fig_exp} (as well
as in Figs. \ref{Fig_pp},\ref{Fig_bm}) is asymmetric in mass due
to the fact that the dilepton decay leads to a multiplication of
the $\rho$-spectral function by $\sim 1/M^3$
since the electromagnetic decay includes a factor $\sim 1/M^4$
from the virtual photon propagator and a factor $\sim M$ from
final state phase space (cf. Ref. \cite{BCM00SIS}). This is in
contrast to the simplifying assumption made in Ref.  \cite{KEK},
where the mass depending branching ratio to dileptons has been
taken as a constant;
the same holds for the $\omega$ mesons. However, accounting for
the $1/M^3$ factor for the $\omega$ and $\rho$ dilepton yields
changes the experimentally fitted curves by a few percent only,
which does not influence the normalization used here (as well as
the conclusions of Ref. \cite{KEK}).

One can see from Fig. \ref{Fig_exp} that the $p C$ data are
roughly reproduced by the calculations within the statistics
achieved experimentally, whereas the $p Cu$ data are
underestimated for $0.55 \le M \le 0.75$ GeV. Thus, the excess of
the dilepton yield in $p Cu$ relative to $p C$ reactions can not
be explained neither by the bare mass scenario (as in Ref.
\cite{KEK}) nor by including in-medium effects such as collisional
broadening and dropping vector masses, which gives only a tiny
enhancement of the dilepton yield (as mentioned above) by less
than a factor of 1.5.

Note, that the calculated yield is higher than the results of the
fit procedure from Ref. \cite{KEK} for $0.55 \le M\le 0.75$ GeV.
That is due to the fact that the calculated $\rho$ contribution is
slightly higher than the fitted one (corrected by the factor
$\sim 1/M^3$).  In the simulations of Ref. \cite{KEK} the assumption
was made, that all vector mesons are produced from primary
nucleon-nucleon collisions at 12 GeV with equal $\rho$ and
$\omega$ production cross sections (in line with the experimental
data on $\rho$, $\omega$ production at 12 GeV/c from Ref.
\cite{Rhoprod}). This leads to an about equal dilepton yield from
$\rho$ and $\omega$ mesons, especially using the assumption about
the instantaneous decay of vector mesons into dileptons, i.e.
without performing the time integration of the  dilepton radiation
during the life time of the mesons (before their nonleptonic
decays such as $\rho^0\to \pi^+\pi^-, \ \omega\to
\pi^+\pi^-\pi^0$).  Note, that the time integration procedure, as
described in Ref. \cite{LKB96},  is important for a correct
evaluation of the dilepton spectra due to the absorption of vector
mesons in the nucleus.

In the resonance-string model the vector mesons at 12 GeV are dominantly
produced in first nucleon-nucleon collisions via  baryon-baryon
($BB$) string excitations and decays with production cross
sections as in Ref. \cite{Rhoprod}. However, $\rho, \omega$ mesons
can be also produced  by secondary meson-baryon ($mB$) collisions
and the $\rho$ mesons additionally by the decay of baryonic
resonances (e.g. $D_{13}(1520)$) formed in primary $BB$ (less frequent) or
secondary $mB$ collisions. The $mB$ and resonance production
mechanisms are more important for heavy nuclei and modify the
$\rho$ mesons spectrum essentially at low invariant mass.
Nevertheless, at 12 GeV -- contrary to the low energy $pA$ reactions --
the resonance dynamics becomes less important since the probability (or
cross section) to form baryonic resonances in $BB$ or $mB$ collisions
(which is high at thresholds) decreases with energy (see, e.g.
\cite{Effe99gam}).
In numbers: in $p C$ collisions $\sim 90$\% of $\rho$ and $\sim 95$\%
of $\omega$ mesons are produced by $BB$ string decays, $\sim 3$\% of
$\rho$ and $\sim 5$\% of $\omega$ by the $mB$ (high energy) string
formation or by secondary $mB$ collisions and $\sim 7$\% of $\rho$ by
the decays of $D_{13}(1520)$ and other baryonic resonances coupled to
the $\rho$. For the heavy $Cu$ target $\sim 70$\% of $\rho$ and $\sim
85$\% of $\omega$ mesons are produced by  $BB$ string decays, $\sim
10$\% of $\rho$ and $\sim 15$\% of $\omega$ by $mB$ interaction and
$\sim 15$\% of $\rho$ by the decays of $D_{13}(1520)$ and other
baryonic resonances.  Note, that the counting of $\omega$ mesons
produced by $mB$ interactions is different from the $\rho$'s due to the
fact that in low energy secondary $mB$ collisions predominantly
baryonic resonances are excited which sequentially might decay to the
nucleon-$\rho$ channel. In the latter case the corresponding $\rho$
mesons are counted in the resonance decay channel inspite of
originating from a $mB$ collision. On the other hand the resonances,
that couple in part strongly to the $\rho$ but not to the $\omega$, can
be also produced in a nucleon-nucleon collision and be reabsorped in
the $RN \rightarrow NN$ reaction by detailed balance.

Note, that the relative numbers quoted above depend on the
number of resonances that are excited during the $pA$ reaction.
These in turn depend on the threshold energy employed for the
string excitation mechanism. However, actual simulations with
reduced/enhanced 'string thresholds' were found to modify the
resulting dilepton spectra by less than 10\%, which is much lower
than the descrepancy (missing yield) in the dilepton spectra below
the $\omega$ peak for $p Cu$ in Fig. \ref{Fig_exp}.
Furthermore, the resonance dynamics essentially modify the dilepton mass
spectra below invariant masses of 0.5 GeV. Indeed, when switching off
all resonances in the transport calculation and producing the meson
directly, the dilepton yield from low mass $\rho$-mesons is modified by
up to a factor of 2. However, summing up all contributing channels the
modification of the total spectrum only amounts to a few percent.  Thus
the systematic uncertainties of the transport calculations in this case
-- when normalized to the experimental $\omega$-peak -- are less than
10\% in the invariant mass range presented in Fig. \ref{Fig_exp}.

To clarify the influence of the nucleon optical potential on dilepton
production at the energy of 12 GeV, additional calculations have been
performed for $pC$ and $pCu$ within the cascade mode, i.e. without any
nucleon optical potential. The results stay the same within the
statistical errorbars achieved in the calculations.

The $\rho$ mesons live for a very short time and radiate dileptons
with high intensity (as compared to the $\omega$ mesons) before
decaying into two pions or being absorbed in the nuclear interior.
The $\omega$ mesons are longer living particles and they can be
absorbed by nucleons in the nucleus, too, before their natural
'death', i.e. 3 pion decay. The $\omega$ absorption violates the
$A^{2/3}$ scaling (where $A$ is the target mass number), i.e. the
scaling of the dilepton rate from $p A$ reactions is expected to
be higher by a factor of $A^{2/3}$ than in $p+p$ collisions
\cite{KEK}. As seen from Fig. \ref{Fig_pp} and Fig. \ref{Fig_bm}
in the calculations the $A^{2/3}$ scaling is practically fulfilled
for the light $C$ target, however, for the heavy $Cu$ target the
dilepton yield from $\omega$ mesons
is  $\sim 30$\% lower as expected from $A^{2/3}$ scaling due
to $\omega$ absorption,
which proceeds dominantly through the $\omega N \rightarrow \pi N$
channel. The absorbed $\omega$'s then can no longer decay to
dileptons which in vacuum would radiate $e^+e^-$ pairs (in their
rest frame) for about 23 fm/c.

\begin{figure}[ht]
\centerline{\psfig{figure=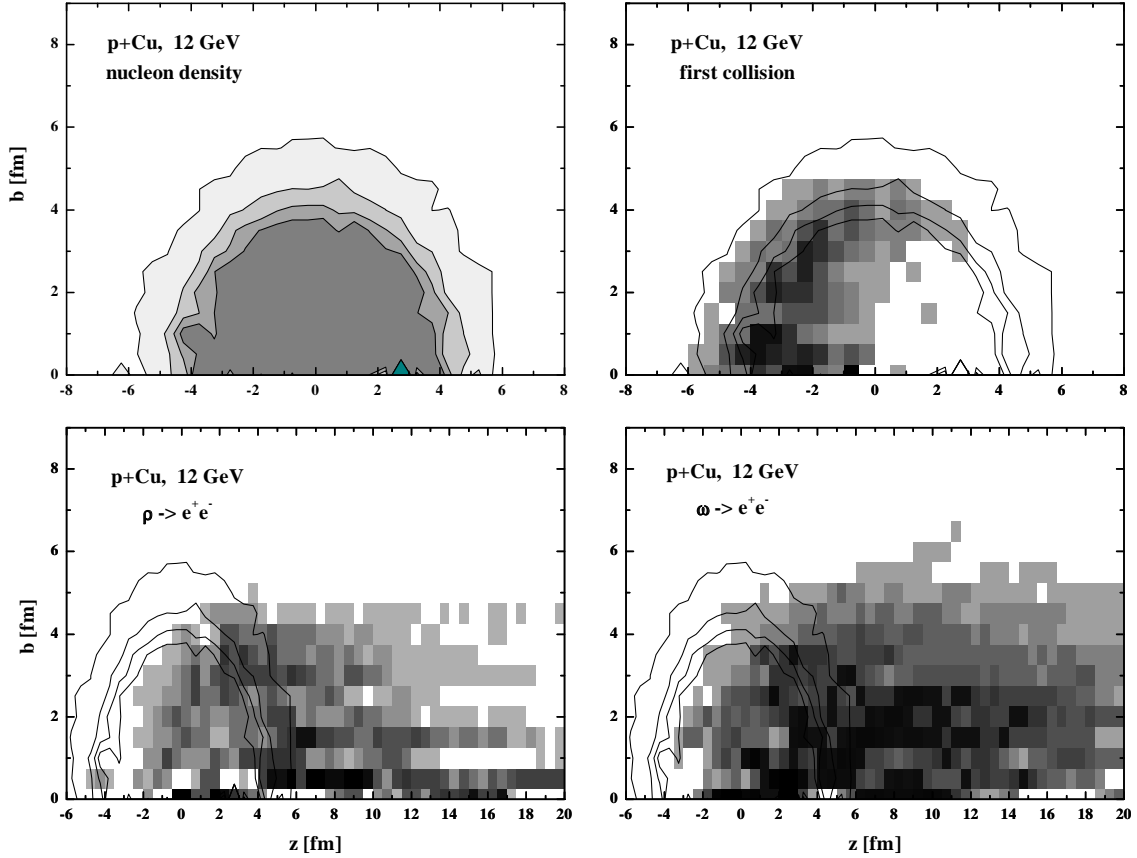,width=15cm}}
\caption{
Upper left part: the average density distribution of a $Cu$-nucleus
at rest in the laboratory; upper right part: the spatial distribution
in the first $pN$ collisions; lower parts:  the spatial distribution
for $\rho$-meson (left) and $\omega$-meson (right) decays
to dileptons. The contour lines correspond to nucleon densities
of 0.1$\rho_0$, 0.4$\rho_0$, 0.6$\rho_0$ and 0.8$\rho_0$,
respectively, and the dark shaded area to $\rho \ge 0.8\rho_0$.}
\label{Fig_bz}
\end{figure}

For a better understanding, why the in-medium effects are so
small, the average density distribution of a $Cu$-nucleus at rest
in the laboratory is shown in Fig. \ref{Fig_bz} (upper left part)
as well as the spatial distribution in the first $pN$ collisions
(upper right part).  Here the spatial distribution ${1\over
b}{dN\over db dz}$ is displayed in cylindrical coordinates
$b=(x^2+y^2)^{1/2}$ and $z$, where $z$ is directed along the beam
axis and the proton is impinging on the nucleus from the left
side. The lower part of Fig. \ref{Fig_bz} displays the spatial
distribution for $\rho$-meson (left part) and $\omega$-meson
(right part) decays to dileptons for the bare mass case.  As
mentioned above, at 12 GeV bombarding energy most of the $\omega$
and $\rho$ mesons -- formed in primary $pN$ collisions -- move
with a high velocity through the target. Contrary to low energy
collisions (cf. Fig. 1 from Ref. \cite{Brat01pA}), most of the
$\rho$'s and $\omega$'s are radiating dileptons and decaying
outside the nucleus; this holds especially for $\omega$ mesons
which have a longer life time such that the $\omega$ spatial
distribution is more elongated in beam direction then the $\rho$.
In numbers: for the $C$ target only $\sim 2.5$\% of $\omega$ and
$\sim 14$\% of $\rho$ mesons decay to dileptons inside the
nucleus; for the $Cu$ target $\sim 7$\% of $\omega$ and $\sim
32$\% of $\rho$ mesons radiate $e^+e^-$ pairs inside. For the
collisional broadening + dropping mass scenario the fraction of
the $\rho$ mesons decaying to lepton pairs inside increases to
$\sim 45$\% for the $Cu$ target due to an enhanced production of
$\rho$ mesons with low masses. Thus the dropping mass and
collisional broadening effects -- which are proportional to the
baryon density -- do not influence the dilepton spectra very much.

\begin{figure}[ht]
\centerline{\psfig{figure=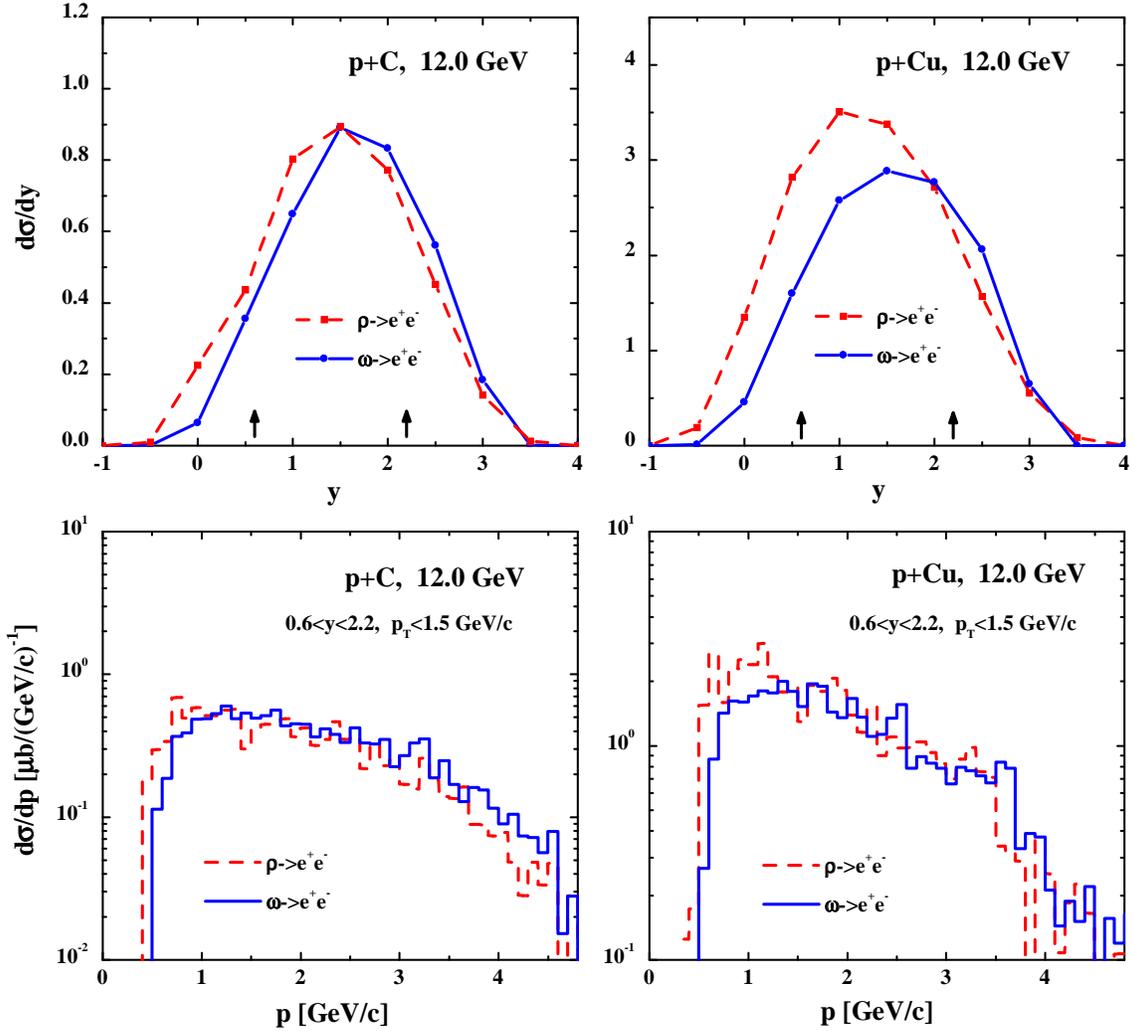,width=15cm}}
\caption{
Upper part: the laboratory rapidity distributions $d\sigma/dy$
for $\rho\to e^+e^-$ (dashed lines) and $\omega\to e^+e^-$ (solid lines)
for $p + C$ (left) and $p + Cu$ (right) at 12 GeV.
The arrows indicate the experimental rapidity window ($0.6\le y\le 2.2$).
Lower part: the momentum distribution $d\sigma/dp$
for $\rho\to e^+e^-$ (dashed lines) and $\omega\to e^+e^-$ (solid lines)
for $p + C$ (left) and $p + Cu$ systems (right) at 12 GeV
calculated with experimental cuts in rapidity ($0.6\le y\le 2.2$)
and $p_T$ ($p_T < 1.5$ GeV/c).}
\label{Fig_yp}
\end{figure}

To illustrate the dynamics of $p A$ reactions at 12 GeV, the
laboratory rapidity distribution $d\sigma/dy$ is presented in Fig.
\ref{Fig_yp} (upper part) for $\rho\to e^+e^-$ (dashed lines) and
$\omega\to e^+e^-$ (solid lines) for $p + C$ (left) and $p + Cu$
(right) at 12 GeV.  The arrows indicate the experimental
rapidity window ($0.6\le y\le 2.2$).  The lower part of Fig. \ref{Fig_yp}
shows the momentum distribution $d\sigma/dp$ for $\rho\to e^+e^-$
(dashed lines) and $\omega\to e^+e^-$ (solid lines) for $p + C$ (left
part) and $p + Cu$ (right part) at 12 GeV calculated with
the experimental cuts on rapidity ($0.6\le y\le 2.2$)
and $p_T < 1.5$ GeV/c.

One can see from the upper part of Fig. \ref{Fig_yp} that the
rapidity distribution for $\rho \to e^+e^-$ and $\omega\to e^+e^-$
are similar for $p C$ collisions; only the $\rho$ distribution is
shifted slightly to the lower rapidity region due to the fact that
the $\rho$ spectral function is broad and $\rho$ mesons can be
produced with low masses, too.  This shift is more pronounced for
the heavy $Cu$ target, where more $\rho$ mesons are formed with
lower masses via  baryonic resonance decays.  Also, the $\omega$
absorption effect for the dilepton yield, as discussed above, is
well seen.

The experimental acceptance selects dilepton events at
mid-rapidity with momentum distributions shown in the lower part
of Fig. \ref{Fig_yp}.  It is seen, that the $\rho$'s and
$\omega$'s are very fast; the corresponding Lorentz
$\gamma$-factors (in the laboratory frame) are in the range $1<
\gamma <6.5$.  One has to recall that the vector mesons produced
by strings can radiate dileptons only after they are formed as
physical particles (not as  pre-resonant or quark states
which have a continuum spectral function).
The formation time $\tau_F$, which denotes the time between the
formation  and fragmentation of the string in the individual
hadron-hadron center-of-mass system is taken to be $\tau_F=0.8$ fm/c
(cf.  \cite{CBRep98,Ehehalt,Bass}). Due to covariance the hadrons with
finite momentum $p$ then are formed at the actual time $\Delta t=\gamma
\tau_F$ (in the laboratory frame).  During this time $\Delta t$ the
pre-formed particles (strings) are moving with high velocities and
start to radiate dileptons only either outside the nucleus or closer to
the nuclear surface (cf.  Fig. \ref{Fig_bz}).  This kinematical effect
suppresses in-medium effects.  It is worth to point out, that for low
energy $p A$ collisions the in-medium effects are expected to be larger
since the vector mesons are formed (and decay to dileptons) more
abundantly inside the nucleus \cite{Brat01pA}.

One might argue that the formation time $\tau_F$ could be substantially
smaller than the value of 0.8 fm/c adopted here (or values of 1-2 fm/c
depending on the hadron-species as used in UrQMD \cite{Bass}). In fact,
when assuming $\tau_F=0$, the amount of in-medium $\rho$ decays for
$p + Cu$ increases to $\sim 50$\% (and to $\sim 65$\% for the
collisional broadening + dropping mass scenario) and to $\sim 12$\% for
the $\omega$ mesons, respectively. In this limit the calculated
dilepton yield indeed becomes comparable to the experimental spectrum
for $p + Cu$.  Furthermore,  the transport calculation for $\tau_F=0$
roughly correspond to the Glauber limit for vector meson production and
absorption in $p + A$ reactions.  However, as shown in the context of
heavy-ion collisions \cite{Ehehalt}, a reduction of $\tau_F$ leads to
an unphysical production multiplicity of mesons (as well as to rapidity
and momentum distributions of baryons and mesons that are incompatible
with the experimental data) since the rescattering rate becomes too
high (some examples for $p + A$ reactions are also shown in Ref.
\cite{EffePhD}).  Furthermore, experimental studies on $\bar p$
production and reabsorption at 12.3 and 17.5 GeV/c in $p + A$ reactions
by the E910 collaboration \cite{pbarAGS} have shown that the formation
time of antiprotons might be even larger than 0.8 fm/c in order to
describe the low amount of $\bar p$ reabsorption in $p + A$ reactions.

In summary: the dilepton production from $p C$ and $p Cu$
collisions at $T_{lab}=12$~GeV has been calculated using a
combined resonance-string transport approach.  It has been found
that the enhanced production of dileptons below the $\omega$ meson
mass pole in $p Cu$ collisions compared to $p C$ -- as observed in
the experiment by the KEK-PS E325 collaboration -- cannot be
explained by collisional broadening nor dropping vector meson masses.
The missing yield cannot be attributed to systematic
uncertainties of the transport approach with respect to the low energy
resonance dynamics as discussed throughout the paper.
It has been argued, furthermore, that such in-mediums effects are
expected to be small due to kinematical reasons, i.e. the dominant part
of vector mesons radiate dileptons and decay outside the nucleus due to
a finite hadron formation time $\Delta t=\gamma\tau_F$.

Furthermore, the dropping mass scenario should be questionable
especially for $\rho$ mesons within the momentum range shown in
Fig. \ref{Fig_yp}  since dispersion relations -- as employed by
Eletsky and Ioffe \cite{ElIof} and Kondratyuk et al.
\cite{Kond98rho} -- lead to repulsive $\rho$ meson potentials for
such high momenta. To clarify such experimental and/or theoretical
inconsistencies high accuracy dilepton spectra within different
momentum bins are urgently needed
as well as independent data on differential vector-meson spectra
from hadronic decays to reduce in addition the theoretical
uncertainties on the cross sections that enter the transport analysis.

\vspace*{5mm}
The author is grateful to U. Mosel for pointing out this problem,
to K. Ozawa for the explanations of the KEK-PS E325 experiment,
and to W. Cassing for valuable discussions and a critical
reading of the manuscript.


\newpage

\phantom{a}\vspace*{3cm}

\end{document}